\documentclass[11pt]{article}
\usepackage{graphicx}
\usepackage{epstopdf}

\title{\bf Student project: Of spinning coins and merging black holes}

\author{Joss Bland-Hawthorn (U. Sydney) \\
\& Rashmi Sudiwala (U. Cardiff)}
\date{}                                           

\begin{document}
\maketitle

\begin{abstract}
For the past decade, the SAIL labs at the University of
Sydney have been challenging students with short
research projects that elucidate basic principles of physics. These include the
development of instruments launched on cubesats,
balloons, on telescopes or placed out in the field.
This experiment is inspired
by the spectacular 2015 discovery of merging black holes
with the Laser Interferometric Gravitational Observatory
(LIGO). Students are profoundly inspired by LIGO, and
for good reason, but it is challenging to construct
a table top demonstration of a gravitational observatory.
Instead we consider chirps which
are remarkable transient phenomena in nature 
involving both frequency and amplitude modulation, as
we can demonstrate with a spinning coin.
In the case of the LIGO event, orbital energy is being
released as gravitational radiation; for the spinning coin,
its spin/orbit energy is being released dissipatively
(sound, heat, air viscosity). Our experiment involves a
simple device to spin a coin remotely. This aids 
repeatability and allows
us to spin the coin within a vacuum chamber to examine
the contribution of air viscosity.\footnote{Experiment movies are posted at
https://www.youtube.com/channel/UCgEqb638Dp2SHvdwTnsaApA .}
\end{abstract}

\section{Introduction}
Many forms of time-varying oscillatory behaviour give rise
to chirps, including speech and music, bat sonar and bird song. Our machines exploit chirp in modern radar systems
and in fibre-based telecommunications.
Chirps also occur ahead of earthquakes, medical seizures and market crashes. 
Intriguingly, chirps occur in mathematics, in particular,
in relation to critical phenomena and power-law divergences,
but also in Weierstrass and Riemann functions.

The mathematics of chirp is subtle because there are 
many forms (e.g. linear, power-law, hyperbolic) and
time-averaged quantities are less useful than for
common band-limited functions (Flandrin 2001). Even so, they have powerful applications in 
signal decomposition and speech simulation, inter alia.

\begin{figure*}
\begin{center}
\includegraphics[scale=0.5]{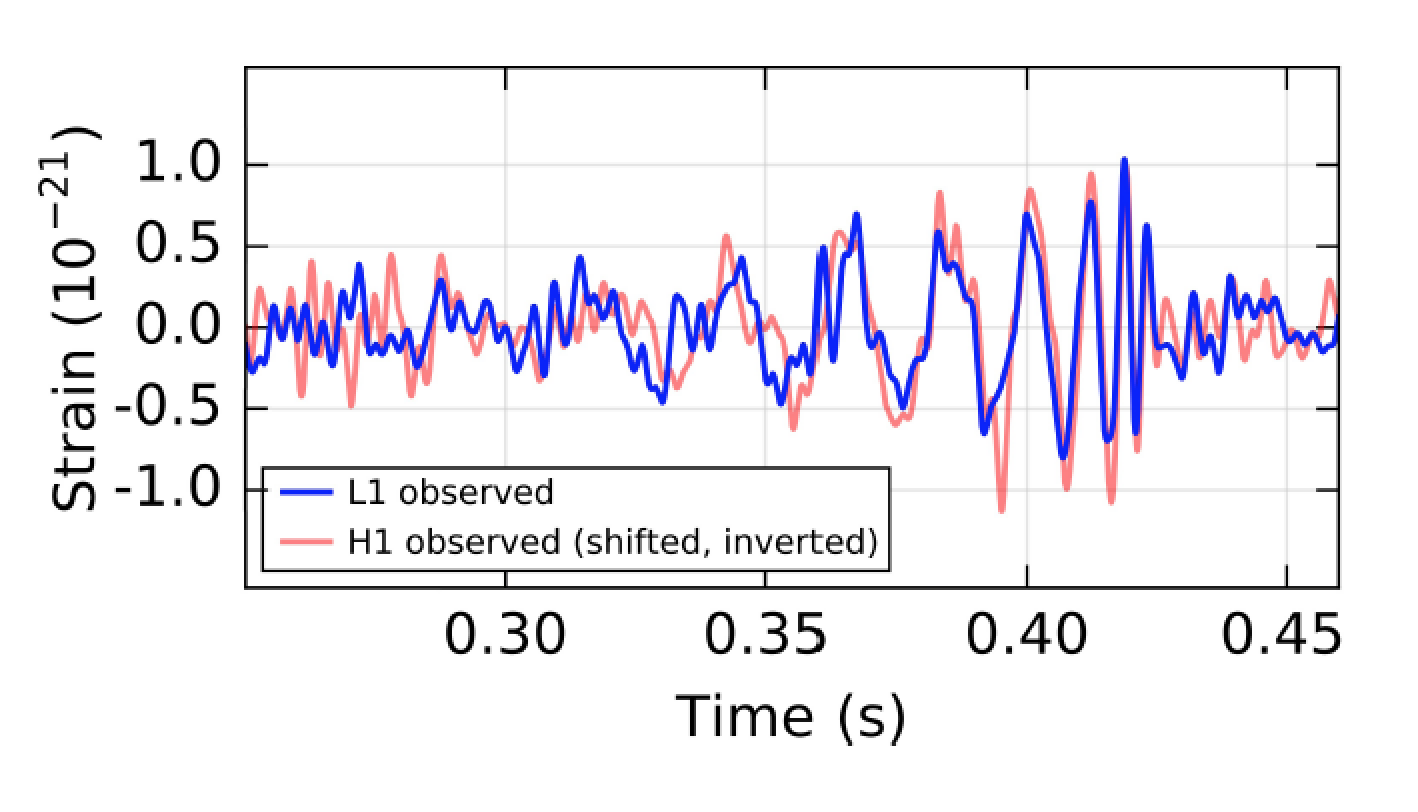} 
\includegraphics[scale=0.5]{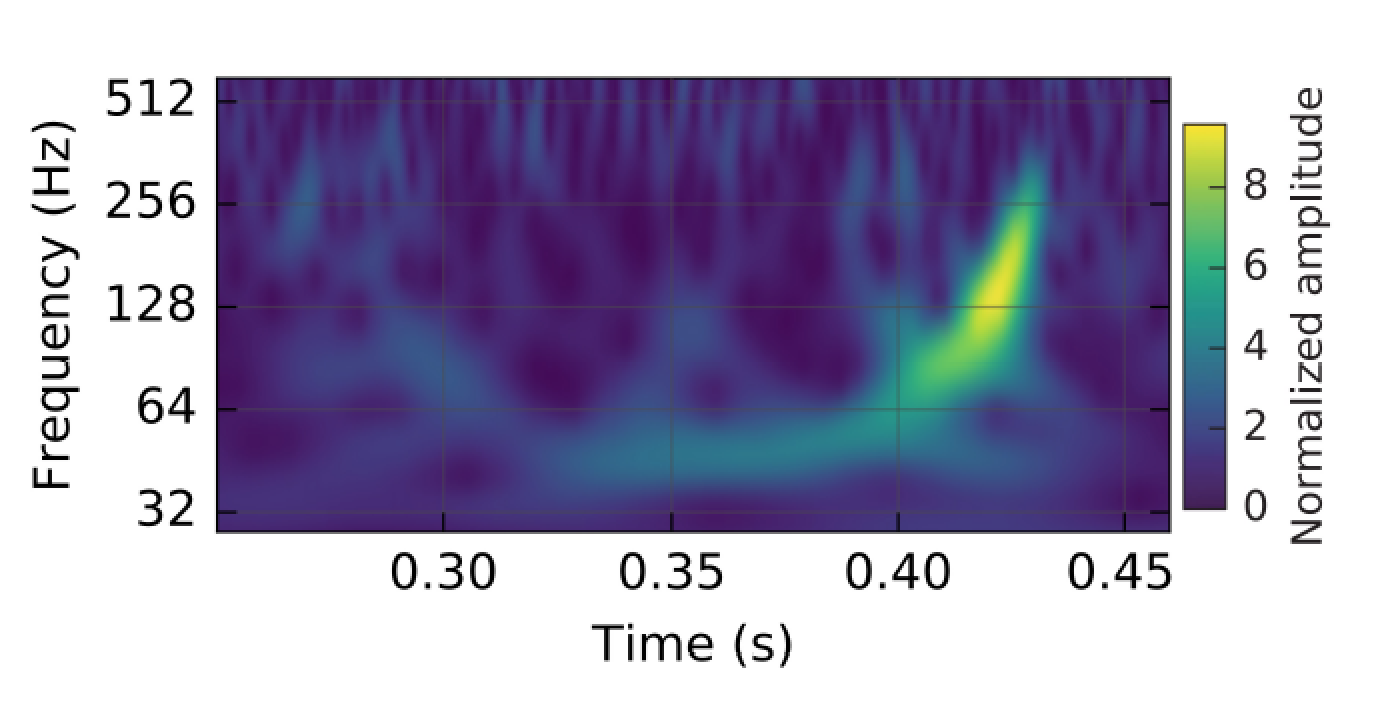} 
\end{center}
\caption{The sensational LIGO discovery of GW150914 published
in Abbott et al (2016$a$). (Top) The results from the
two LIGO observatories are overlaid. (Bottom) The
L1 data presented as the frequency of the gravitational
radiation with time. The strain amplitude is colour
coded.
}
\label{f:LIGO}
\end{figure*}

\section{Chirp signal from the LIGO discovery event}

A good student can quickly derive and plot the LIGO
chirp for merging black holes using classical physics with help from
Mathur et al (2016) and Abbott et al (2016$b$). To a small factor,
much of what is observed can be understood in terms
of Newtonian physics as long as one
compels the binary system to ``radiate'' its inertia
through the quadropole moment. A useful analogy is 
with dipole
radiation from a radio beacon due to the changing 
electric or magnetic dipole moment of the antenna.
A fundamental difference
is that the frequency of gravitational radiation is
twice the frequency of rotation since the quadrupole 
moment of equal or unequal masses is invariant under a 
rotation of $\pi$.

\begin{figure*}
\begin{center}
\includegraphics[scale=0.5]{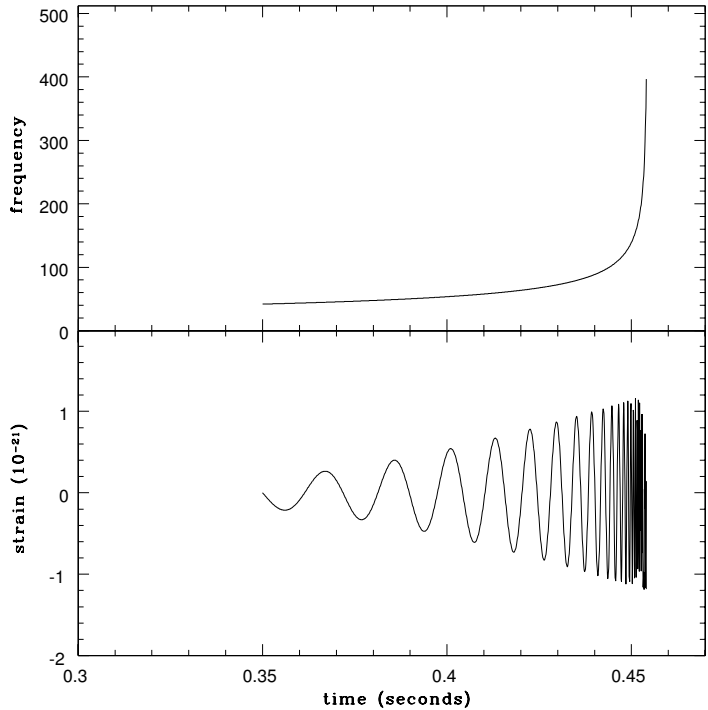} 
\end{center}
\caption{Simple model to compare with the LIGO discovery of the black hole binary merger in Fig.~\ref{f:LIGO}. (Top) The solution to Eq.~\ref{e:freq} after 
integration where the initial condition is
set by the data at $t=0.35$ sec in Fig.~\ref{f:LIGO}; (Bottom) The predicted strain amplitude (normalized to the second 
derivative of the quadrupole moment). The simple formulae
use only Newtonian dynamics (with enforced quadrupole
emission) and do not accommodate any form of damping (ring down).
}
\label{f:mathur}
\end{figure*}

If two black holes $M$ and $m$ are separated by $R+r$
such that $Mr=mR$, Kepler's laws lead us to
\begin{equation}
\omega^2 = {{G(M+m)}\over{(R+r)^3}}
\end{equation}
for circular orbits; eccentric orbits quickly circularize
in any event. The total energy of the system (orbital$+$gravitational) is
\begin{eqnarray}
E_{\rm tot} &=& \frac{1}{2} {{GMm}\over{(R+r)}} \\
            &=& \frac{1}{2} {{G^{2/3}\omega^{2/3}Mm}\over{(M+m)^{1/3}}}.
\end{eqnarray}
The moment of inertia of the system about the centre of mass is 
\begin{equation}
I = {{Mm}\over{M+m}}(r+R)^2
\end{equation}
which is essentially the mass quadrupole of the binary system. Since the power radiated must be proportional to the
square of the amplitude, simple dimensional analysis
(Mathur et al 2016) leads
to
\begin{equation}
P_{\rm rad} \propto {{GI^2\omega^6}\over{c^5}}
\end{equation}
where the dimensionless constant of proportionality
(32/5) requires some general relativity (metric tensors). The radiated energy
must be equal to the change in total energy such that
$P_{\rm rad} = -dE_{\rm tot}/dt$ which brings in the
time derivative $\dot{\omega}$. By substitution, it follows
that
\begin{equation}
{{(Mm)^{3/5}}\over{(M+m)^{1/5}}}={{c^3}\over{G}}({{5}\over{96}}\omega^{-11/3}\dot{\omega})^{3/5}.
\end{equation}
The LHS of this equation is referred to as the chirp
mass ${\cal M}$: this has the form of a mass as is seen by setting
$m=M$. The summed masses $M+m$ are always greater than or
equal to $2{\cal M}$. If we insert the frequency $f$ ($=\omega/\pi$) of the observed gravitational radiation into the above equation then
\begin{equation}
{\cal M}={{c^3}\over{G}}({{5}\over{96}} \pi^{-8/3} f^{-11/3}\dot{f})^{3/5}
\label{e:freq}
\end{equation}
which describes a rapid chirp through the modulated
frequency dependence. The increase in amplitude comes
from the dependence of $P_{\rm rad}$ on the frequency $f$.
As Mathur et al (2016) observe, this is the precise equation
quoted in the LIGO discovery paper (Abbott et al 2016$a$).
As Abbott et al (2016$b$) show, Eq.~\ref{e:freq} can be 
integrated with respect to time to allow $f$ to be
expressed as an explicit function of time. This is plotted
in Fig.~\ref{f:mathur} along with a simple prediction
for the strain amplitude (Abbott et al 2016$b$).

It is interesting to note the paper by Regge \& Wheeler (1957)
on the stability of the Schwarzschild singularity. After 
ring down, it seems likely that the merged black hole continues
to oscillate forever but with small perturbations. Maybe one
day our descendants will be able to detect the low-amplitude
oscillatory behaviour of all black holes. By then, we
will have learnt to ``image'' gravitational waves from
the Universe, to measure the cosmic background and to form
spherical harmonics in both polarisation states. There is no
end to physics.

\section{Chirp signal from a spinning coin}

Spinning coins are a nice demonstration of a 
finite-time singularity.
As van den Engh et al (2000) observe,
``the familiar shuddering motions of spinning coins
as they come to rest are not at all intuitive.''


Assume a coin with a rounded (bevelled) rim is rolling on its edge over a flat surface without slipping and
with an angular velocity $\omega_r$. Once the coin slows
enough, it starts to tip over and wobble about
a vertical axis perpendicular to the flat surface. 
The coin's motion is now a combination of rolling 
($\omega_r$)
and precession defined by an angular velocity 
($\omega_p$) about 
the vertical axis, i.e. the precession of the disk rolling
vector about the vertical axis. If the coin's tilt
angle is $\alpha$ with respect to the surface, the
circle drawn by the rim of the precessing disk is smaller than the coin's perimeter. Thus for $\alpha < \pi/2$
\begin{equation}
\omega_r = \omega_p(\sec\alpha-1)
\label{e:spin}
\end{equation}
As $\alpha$ gets smaller, $\omega_r \rightarrow 0$ and
the coin settles into
a rapid sequence of wobbles ($\omega_p$ rising) before it stops abruptly for reasons that are not captured
in Eq.~\ref{e:spin}. 
Different authors have contrasting views on what causes
the coin wobble to halt rather than rise asymptotically
to infinite frequency. For example, Moffatt (2000) argues
that air viscosity is dominant, although Bildsten (2002)
finds that contact forces are more likely to dominate.
We discuss these effects below.


\begin{figure}
\begin{center}
\includegraphics[scale=0.7]{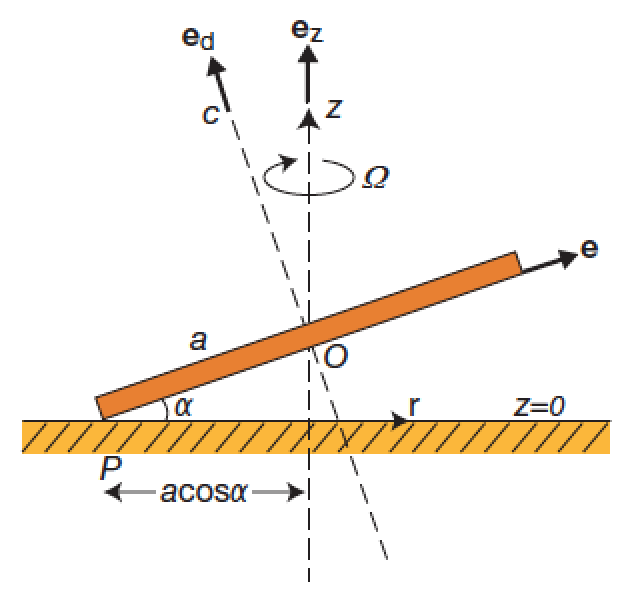} 
\end{center}
\label{f:moffatt}
\caption{Geometry of a spinning coin (Moffatt 2000):
$\Omega$ is the total angular velocity (rolling$+$wobble)
of the spinning disk about the vertical axis. This is
solved in terms of the components about $\vec{e}_d$ (rolling) and $\vec{e}$ (wobble).
}
\end{figure}

Moffatt (2000) has presented a dynamical analysis and
we use his notation in Fig.~\ref{f:moffatt}. He considers
the total angular momentum about a vertical axis as the
sum of the spinning disk about its spin axis, and the
wobbling disk about an axis in the plane of the disk.
(The line {\it OP} defines an instantaneous axis of
rotation as the disk wobbles about it.)
The angular velocity of the disk $\vec{\omega}$ acting about
the vector $\vec{e}$ is given by
\begin{equation}
\vec{\omega} = \Omega\cos\alpha\: \vec{e_d} - \Omega\: \vec{e_z}
\end{equation}
Thus $\omega=\vec{\omega}.\vec{e}=-\Omega\sin\alpha$
and $\omega=\vec{\omega}.\vec{e_d}=0$ as expected.
The total
angular momentum of the disk is $\vec{h}=A\omega\vec{e}$
where $A=\frac{1}{4}Ma^2$ is the moment of inertia of a disk of mass $M$ about its diameter $2a$.
From Euler's equation, the gravitational torque $\vec{G}$
about the contact point is given by
\begin{eqnarray}
\vec{G} &=& \vec{\dot{h}}\:\: =\:\: \vec{\Omega} \times \vec{h} \\
  &=& Mga\: \vec{e_z} \times \vec{e}
\end{eqnarray}
which leads to
\begin{eqnarray}
\Omega^2 &=& 4g/a\sin\alpha \\
\Omega^2 &\approx& 4g/a\alpha {\rm\:\:\: for\: small}\: \alpha
\end{eqnarray}
where $g$ (=980 cm$^2$ s$^{-1}$) is the acceleration due to Earth's gravity. 

The finite-time singularity is now clear: as $\alpha$ decreases
due to dissipation, the coin wobble frequency increases
asymptotically since $\Omega = 1/\sqrt{\alpha}$ to infinite
frequency. But since nature abhors a singularity, this of
course never happens, at least not in this Universe.\footnote{There are youtube videos of Euler disks asymptoting to wobble frequencies 
in excess of 100 Hz which underscores what is meant by a finite-time singularity when seen.}

On energetic grounds,
when the coin is held
vertically or simply balanced on its edge, it has 
excess potential energy with respect to the rest position 
when lying on its side. The impulse we give to spin the
coin defines its maximum (potential $+$ kinetic) energy.
As the coin tips over, the spin is communicated to 
the rolling and precessing motion as we have described:
\begin{eqnarray}
E &=& {\rm PE\: +\: KE} \\ 
  &=& Mga\sin\alpha + \frac{1}{2} Mga\sin\alpha \\
  &=& \frac{3}{2} Mga\:\alpha  {\rm\:\:\: for\: small}\: \alpha
\end{eqnarray}

Moffatt (2000) presents a case for damping the asymptotic
behaviour through air viscosity. First consider the
impact of pushing air out of the way over a distance $a$
in a time given by the wobble rate. The 
viscous dissipation of energy over the displaced
volume below the coin is $\nu \approx \pi\mu g a^2/\alpha^2$ where $\mu$ is the air's viscosity; note that $\nu \rightarrow \infty$ as $\alpha\rightarrow 0$. Finally, we
equate the change in the total energy with the energy
lost to viscosity such that
\begin{equation}
\frac{3}{2} Mga\:\dot{\alpha} = -\pi\mu g a^2/\alpha^2
\end{equation}
which integrates to 
\begin{equation}
\alpha^3 = 2\pi \mu a (t_0 - t)/M.
\end{equation}
This has the desired form that $\alpha$ goes to zero in
finite time $t_0$ but at the cost of the angular
velocity $\Omega \approx (t_0-t)^{-1/6}$ becoming singular!
Moffatt (2000) observes that the size of the vertical 
acceleration $\vert a \ddot{\alpha} \vert$ cannot exceed
$g$ such that the above formula breaks down at time 
$\tau$ before $t_0$ such that
\begin{equation}
\tau \approx (2a/9g)^{3/5}(2\pi \mu a/M)^{1/5}
\end{equation}
Moffatt carried out experiments with settling disks and
found that the settle times are about right within
20\% of this approximation. Thus he believed that
air viscosity was dominant in suppressing the singularity.
Interestingly, Moffatt was able to achieve $\Omega
\approx 500$ Hz with a commercially available ``Euler disk'',
i.e. 4 orders of magnitude change in the wobble
frequency, demonstrating beautifully the notion of
a quasi-singularity in finite time. Bildsten (2002)
constructed a more sophisticated model for the air
boundary and derived $\Omega \approx (t_0-t)^{-2/9}$,
in better agreement with  spinning disk observations
($\Omega\approx 20-70$ Hz) by McDonald \&
McDonald (2000).

Several authors have questioned Moffatt's viscosity argument.
van den Engh et al (2000)
notes there are 20\% differences observed between vacuum
and air at early times of spinning upright, but not during
the settling phase. They find that
different spinning structures (e.g. annular rings, convex lids) have similar settling times thereby
downplaying the role of viscosity, turbulence or other
kinds of air flow dynamics. They emphasize the importance of
surface friction, for example, by spinning an Euler disk
on a table top rather than a smooth surface, and finding
that it comes to a halt quickly. They also identify a
new kind of ``rolling friction'' when rough-edged coins are spun on a table top; see also Petrie et al (2002) and 
Bildsten (2002).

\begin{figure}
\begin{center}
\includegraphics[scale=0.1]{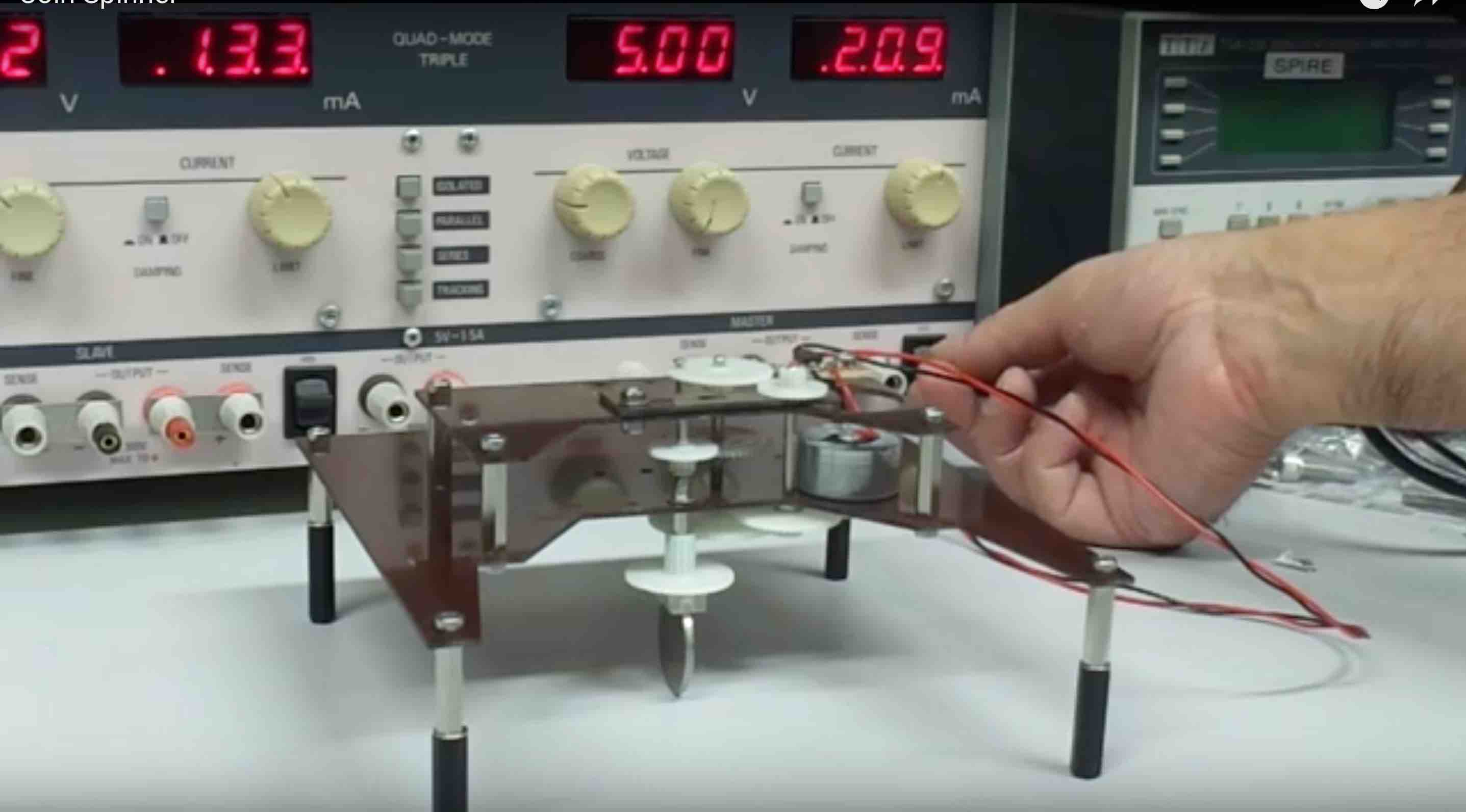} 
\end{center}
\caption{Automated machine to spin a coin used in our
experiments below. See the
youtube video
https://www.youtube.com/watch?v=uO62dLAcr88 .
}
\label{f:rashmi}
\end{figure}

\begin{figure}
\begin{center}
\includegraphics[scale=0.4]{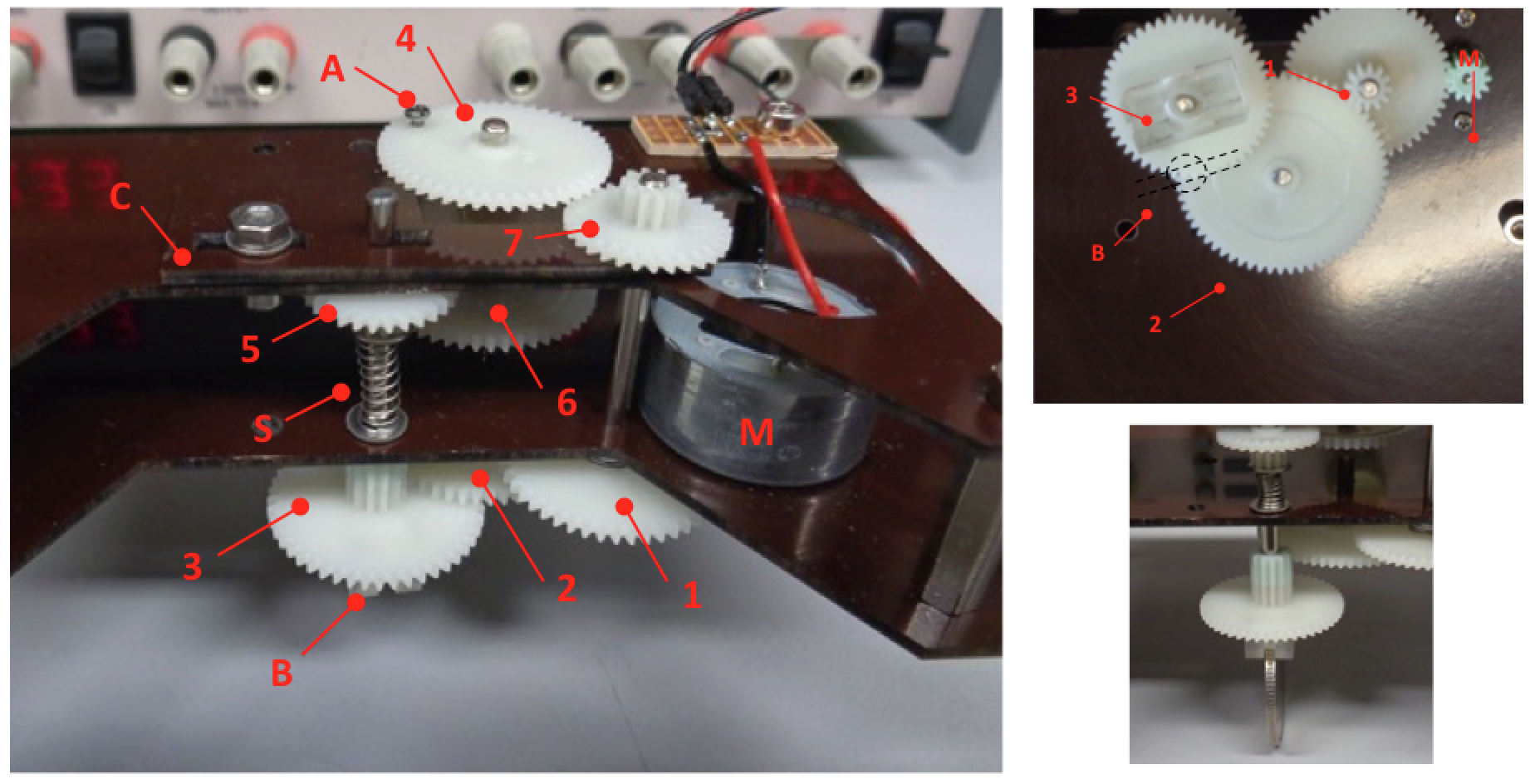} 
\end{center}
\caption{The coin spinner mechanism is operated from a single motor M. The pinion on M turns gear 1 on shaft 1 [12:48 ratio]; gear 1 turns gear 2 on shaft 2 [12:60 ratio]; and gear 2 turns gear 3 on shaft 3 [60:12 ratio]. Gear 3 can be depressed against spring S to engage and hold a coin in block B, and is locked in this position by sliding plate C to the right. As shaft 2 rotates clockwise, pin A on gear 4 pushes plate C back towards the left allowing shaft 3 to spring up, thereby releasing the spinning coin. Wheels 5 and 6 form stops to prevent shaft 3 from being depressed too far. Wheel 7 simply holds shaft 1 in position. The coin holder block B is machined from Perspex: 16x8x6 mm (LWH) with a narrow, 2 mm deep slot running its length and a 5 mm diameter hole through hole; the slot allows easy placement of the coin while the hole helps locate it centrally under the shaft. The mechanism plates are machined from high gloss, low friction MDF drill entry/exit backing boards (e.g. part number 700-015-1 from MegaUK.com).
}
\label{f:spinner}
\end{figure}

\section{Experiment: How to spin a coin with no hands}
van den Engh et al (2002) provide the interesting fact
that the Dutch 2.5-guilder coin has magnetic properties
allowing it to be spun at a precise frequency on a magnetic
stirrer. Being Australians, thence far removed from
guilders and stirrers, we decided to build a device 
that could spin an Australian 10 cent coin;
see Fig.~\ref{f:rashmi} and Fig.~\ref{f:spinner} for
details of the mechanism. We present
a demonstration of the device in a youtube video
\footnote{https://www.youtube.com/watch?v=uO62dLAcr88}. 
It needed to
be compact enough to fit into a vacuum chamber at the
University of Sydney (15 cm span).

First, we filmed the spinning coin in slow motion in
air (240 fps) 
using an iPhone 6. The sound was also recorded with
an iPhone using the Recorder Plus app. See the movie
referred to under Footnote 1. These data can
be read into Matlab or Python and analysed.

\begin{figure}
\begin{center}
\includegraphics[scale=0.4]{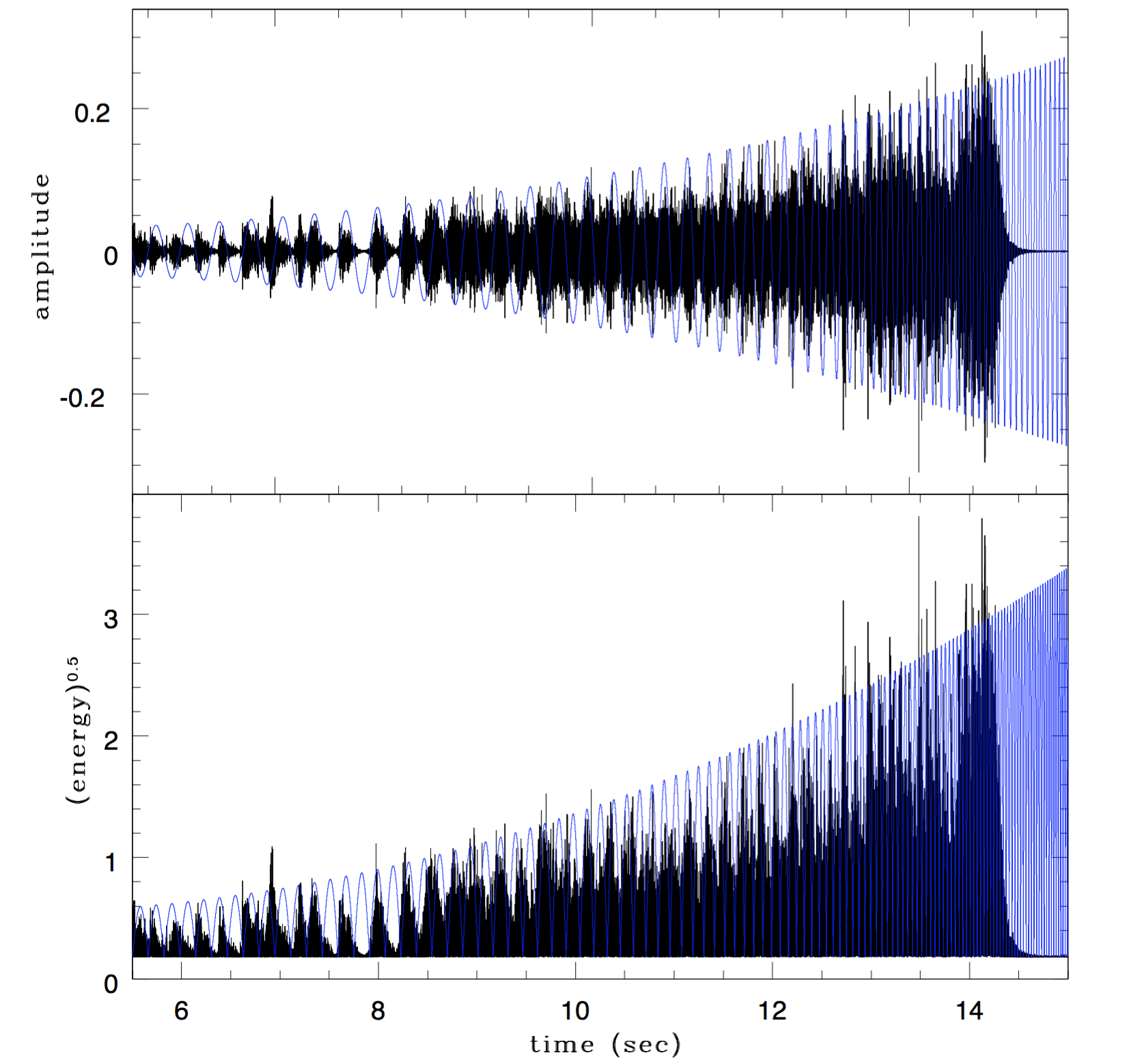} 
\end{center}
\caption{(Top) Sound recording shown as amplitude vs time of a spinning coin. As the contact point spins
into view, the sound rises (rising sine)
before fading as it recedes (falling sine). The rise
and fall are approximately represented by a chirped
sine wave - see lower figure.
(Bottom) Square root of the squared
amplitude to emphasize the energy peaks of the sound
packets. The instantaneous frequency spacing of these 
peaks is shown in Fig.~\ref{f:freq}. The peaks and
nulls are supposed to line up with the sine wave peaks
but this is not always the case due to imperfections
in our set up.
}
\label{f:chirp}
\end{figure}

In our experiment, we record the sound of the spinning
coin in air.\footnote{We considered experiments based on reflected light but these are more problematic.}
Each time the spin motion is directed
towards the microphone, the burst of sound generates a
sound packet that we resolve with the iPhone. The far
side sound is partly absorbed by the coin. Thus during
the coin spinning, we see a succession of sound packets
that become more compressed (increasing instantaneous 
wobble frequency) and the amplitude increases
dramatically. In Fig.~\ref{f:chirp}, the rising amplitude
is approximately in phase with the rising sine wave.
The lower plot shows how the 
peaks and nulls both line up with the positive peaks
of the squared wave. This behaviour is specific to a 
sound experiment. If we had used a 
laser pointer illuminating a spinning coin,
for example, the reflected light leads to an oscillating
pattern at twice the frequency and in phase with
all, rather than alternating, peaks of the squared wave.


\begin{figure}
\begin{center}
\includegraphics[scale=0.3]{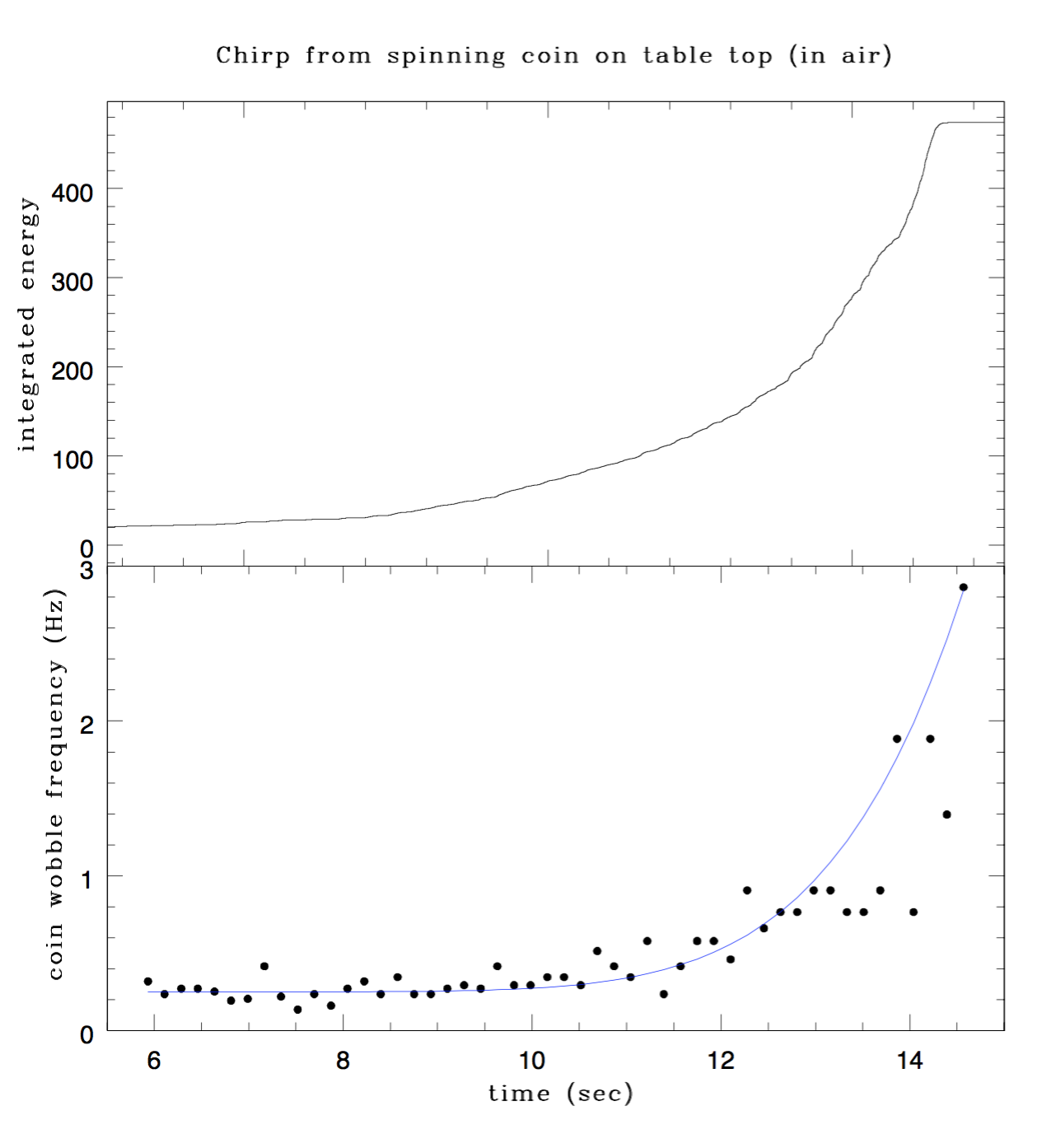} 
\end{center}
\caption{(Top) The integrated energy in Fig.~\ref{f:chirp}
with time. (Bottom) The increase in the instantaneous
frequency in Fig.~\ref{f:chirp} with time.
}
\label{f:freq}
\end{figure}

\begin{figure}
\begin{center}
\includegraphics[scale=0.25]{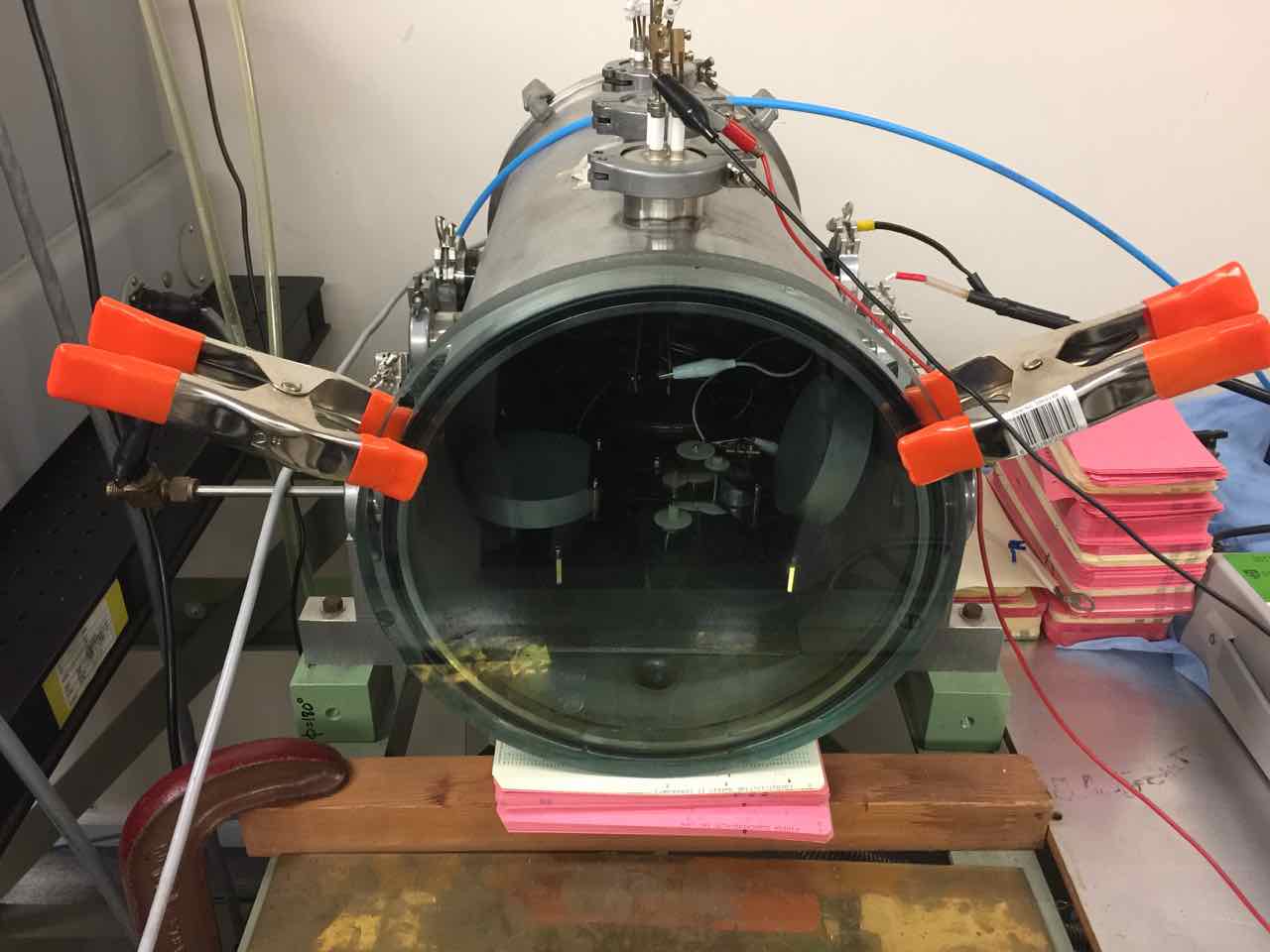} 
\end{center}
\caption{The automated coin spinner loaded into the 
vacuum chamber in the Space Lab at the School of Physics,
University of Sydney.
}
\label{f:chamber}
\end{figure}

\section{Vacuum chamber experiment}

We were able to obtain slow-motion movies of a coin
spinning in a vacuum; the link to the movie is given 
in Footnote 1. The chamber has vacuum seals that allow
for electronic control of a remote device, as shown.
The chamber can be evacuated to 1/7000 of
an Earth's atmosphere in about 2 mins. This allows for
rapid cycling of the chamber for repeating spin tests.
We find that the reduced air resistance allows for the
coin to spin upright for about 40\% longer under highly
repeatable conditions, but the final settle phase is
not changed compared to atmospheric conditions. We do
not refer to this as the ``ring down'' phase by 
analogy with LIGO because the sound amplitude is still
increasing.

\section{Comparing chirps}

We have considered two problems in analytic dynamics
which exemplify the remarkable phenomenon of chirp.
The advantage of the coin experiment is that the data 
are easily gathered from repeated experiment, and easily
analysed. To keep our report to a manageable length,
we have refrained from including the
power spectrum analysis of either the LIGO or the
coin data (e.g. Flandrin 2000).
While it is possible for a student to manipulate
processed LIGO data (e.g. Fig.~\ref{f:LIGO}), 
the analysis of raw data is far more involved and not
broached here. For the spinning coin experiment, a refinement would be to attach vibration sensors (and
maybe heat sensors) to the
metal base to record all contact. Another improvement is
to use a magnetic disk spun by a magnetic stirrer because
the torque impulse can be specified accurately.
It will be possible
to temporarily resolve many forms of contact, slippage,
etc. This will keep the student engaged over a month
or so. After that, they will be fired up for original
research based on more challenging experiments.

\bigskip
The SAIL labs are currently supported by an ARC 
Laureate Fellowship (2014-2019)
awarded to JBH and are directed by Sergio Leon-Saval.
We thank Alex Frazis for asking for something new to
do for his first year project. We are also grateful
to Joe Khachan and Joe Builth-Williams for access to the Space Lab vacuum chamber.

\bigskip\noindent
{\bf References}

\smallskip\parindent 0.1cm
Abbott, B.P. et al 2016$a$, Phys. Rev. Lett. 116, 1102

Abbott, B.P. et al 2016$b$, arXiv:1608.01940

Bildsten, L. 2002, Phys. Rev. E. 22, 056309

Flandrin, P. 2001, Proc. SPIE 4391, Wavelet Applications VIII, 161

Mathur, H., Jones-Smith, K. \& Lowenstein, A. 2016,
arXiv:1609.09349 

McDonald, A.J. \& McDonald, K.T. 2000,
adsabs.harvard.edu/abs/2000physics...8227M

Moffatt, H.W. 2000, Nature, 404, 833

Petrie, D., Hunt, J.L. \& Gray, C.G. 2002, Am. J. Phys. 70, 1025 

Regge, T. \& Wheeler, J.A. 1957, Phys. Rev. 108, 1063

van den Engh, G., Nelson, P. \& Roach, J. 2000, Nature, 408, 540

\end{document}